\documentclass[twocolumn,10pt]{article}
\usepackage{psfig,epsfig,subfigure}
\author{Sorav Bansal \hspace{4mm} Mary Baker\\ {\em Stanford University}\\ {\em Stanford, CA 94305,
USA}\\
{\tt sbansal@stanford.edu, mgbaker@cs.stanford.edu}
} 

\begin{document} 
\title{Observation-based Cooperation Enforcement in Ad hoc Networks}
\date{}
\maketitle
\thispagestyle{empty}

\begin{abstract}
Ad hoc networks rely on the cooperation of the nodes participating
in the network to forward packets for each other. A node may decide not to
cooperate to save its resources while still {\em using} the network to 
relay its traffic. If too many nodes exhibit this behavior,
network performance degrades and cooperating nodes may find themselves
unfairly loaded.
Most previous efforts to counter this behavior
(\cite{BB01},\cite{BB02},\cite{BB03},\cite{MGLB00})
have relied
on further cooperation between nodes to exchange reputation information
about other nodes. If a node observes another node not participating
correctly, it reports this observation to other nodes who then take
action to avoid being affected and potentially
punish the bad node by refusing to forward its traffic.  Unfortunately,
such second-hand reputation information is subject
to false accusations and requires maintaining trust relationships with
other nodes.  The objective of OCEAN is to avoid this trust-management
machinery and see how far we can get simply by using direct first-hand
observations of other nodes' behavior.
We find that, in many scenarios, OCEAN can do as well as, or even better than,
schemes requiring second-hand reputation exchanges. This encouraging
result could possibly help obviate solutions requiring trust-management
for some contexts.
\end{abstract}

\section{Introduction}

An {\em ad hoc network} is a group of wireless mobile computers (or
nodes), in which nodes cooperate by forwarding packets for each other
to allow them to communicate beyond direct wireless transmission range.
Ad hoc networks require  no centralized administration or fixed network
infrastructure such as base stations or access points, and
can be quickly and inexpensively set up as needed. They can be used in
scenarios in which no infrastructure exists, or in which the existing
infrastructure does not meet application requirements for reasons
such as security or cost.

So far, applications of mobile ad hoc networks have been envisioned
mainly for crisis solutions (e.g., in the battlefield or in
rescue operations). In these applications, all the nodes of the
network belong to a single authority (e.g. a single military
unit or a rescue team) and have a common goal. For this reason, the
nodes are naturally motivated to cooperate.

With the progress of technology, however, it is becoming possible to deploy mobile
ad hoc networks for civilian applications as well. Examples include networks
of cars, provision of communication facilities in remote areas, and exploiting
the density in urban areas of existing nodes such as cellular telephones to
offload or otherwise avoid using base stations.
In these networks, the nodes may not
belong to a single authority and they do not pursue a common goal.
In addition, these networks could be larger, have a longer lifetime,
and they could be completely {\em self-organizing}, meaning that the network
could be run solely by the operation of the end-users. In such networks,
there is no good reason to assume that the nodes cooperate. Indeed,
the contrary is true: some nodes may be disruptive and others may
attempt to save resources (e.g. battery power,
memory, CPU cycles) through ``selfish'' behavior.

In this paper we describe OCEAN, in which we focus on the robustness
of packet forwarding:
maintaining the overall packet throughput of an ad hoc network in the face of
nodes that misbehave at the routing layer.
We concentrate our efforts at the routing layer and do not
attempt to address attacks at lower layers (eg. jamming
the network channel) or passive attacks like eaves-dropping.  We also
do not deal (much) with issues like node authentication, securing routes, or
message encryption.  Instead, secure routing protocols
\cite{HJP02},\cite{HPJ01},\cite{PH02},\cite{Roy02} are designed to combat those threats.
OCEAN addresses an orthogonal issue -- the encouragement
of proper routing participation -- and can be used in addition to secure routing
protocols to respond to a more complete threat model.  We also do not consider
the collusion of nodes in a network, but merely the individual bad behavior
of nodes.

OCEAN considers two types of routing misbehavior.  The first, which we call
\emph{misleading}, is that a node may respond
positively to route requests but then fail to forward the actual packets, misleading
other nodes into unsuccessfully sending their traffic through it.
Previous approaches at mitigating misleading routing misbehavior \cite{BB01},\cite{BB02},\cite{BB03},\cite{MGLB00}
require nodes in the network to exchange
reputation information about other nodes.  If a node observes another node participating
incorrectly, it reports this observation to other nodes who then take
action to avoid being affected by the misbehavior and perhaps even punish the
node by refusing to forward its traffic.

While these schemes have
proved effective, exchanging second-hand reputation information opens up a new
vulnerability, since nodes may falsely accuse other nodes of misbehaving.  Making
a decision about whether to believe an accusation
requires authenticating and trusting the accusing node.
Such trust maintenance could be
performed offline or could be bootstrapped during network operations. In
the former case, the network requires a priori trust relationships
that may not be practical in truly ad hoc networks.  In the latter case,
bootstrapping trust relationships in ad hoc networks involves
significant complexity and risk and may not be reasonable for a very dynamic
or short-lived network.

OCEAN's approach to this problem is to disallow any second-hand reputation
exchanges. Instead, a node makes routing decisions based solely on direct observations
of its neighboring nodes' exchanges with it.  This eliminates most trust management
complexity, albeit
at a cost of less information with which to make decisions about node behavior.
To our surprise, though, we find that less information does not necessarily
mean less performance.  Using OCEAN we are able to achieve
performance (packet throughput) comparable to that of approaches requiring second-hand 
information exchanges.
On the positive side, OCEAN achieves this while being less complex and
less vulnerable to false accusations.  On the negative side, OCEAN is more sensitive
to some parameter settings and does not punish misbehaving nodes as severely as
systems using full-blown reputation information.

The second type of routing misbehavior we address, which we call \emph{selfish},
is that a node may not even respond to route requests but may nonetheless send
its own traffic through the network, unfairly preserving its resources while exploiting
others'.  This type of misbehavior can be hard to detect, except through observing
the actual data forwarding behavior 
of neighboring nodes.  In OCEAN, we again focus on detecting
this misbehavior with only direct observations of neighboring nodes.
We address the problem using simple, light-weight economic methods that, while
not guaranteed to be fair, nonetheless generally result in reasonable performance.

Section~\ref{sec:relwork} describes recent work related to the problem
of managing misbehavior at the routing layer in ad hoc networks.
Section~\ref{sec:overview} presents an overview of the modules used in OCEAN
to mitigate such routing misbehavior. Section~\ref{sec:selfish} describes
a scheme to deter selfish behavior.
Section~\ref{sec:simresults} discusses the simulation results we obtained
and compares OCEAN to reputation-based approaches.  Section~\ref{sec:discussion}
gives more detail about node authentication issues in OCEAN.  Finally,
Section~\ref{sec:conclusion} concludes.

\section{Related Work}
\label{sec:relwork}
Recently, the problem of security and cooperation enforcement has
received considerable attention by researchers in the ad hoc
network community.

The problem of securing the routing layer using
cryptographically secure messages is addressed by
Hu et al. \cite{HJP02} \cite{HPJ01}, Papadimitratos and Haas \cite{PH02}, and Sanzgiri et al. \cite{Roy02}.
Schemes to handle authentication in
ad hoc networks assuming trusted Certificate Authorities
have been proposed by Zhou and Haas \cite{ZH99}, and Kong et al.  \cite{KZL01}.
Hubaux et al. \cite{HBC01} employ a self-organized PGP-based
scheme to authenticate nodes using chains of certificates and transitivity
of trust. Stajano and Anderson \cite{SA99} authenticate
users by `imprinting,'
in analogy to ducklings acknowledging the first moving subject
they see as their mother.  In OCEAN, we do not attempt to secure the routing
layer, although our techniques may be used in conjunction with many
secure routing protocols to increase performance and robustness.

In contrast to securing the routing layer of ad hoc networks, some researchers have
also focused on simply detecting and reporting misleading routing misbehavior.
{\em Watchdog and Pathrater} \cite{MGLB00} use observation-based
techniques to detect misbehaving nodes
and report observed misbehavior back to the source of the traffic.
{\em Pathrater} manages trust and route selection based
on these reports.  This allows nodes to choose better paths along which to
route their traffic by routing around the misbehaving nodes.  However, the
scheme does not punish malicious nodes; instead, they are relieved
of their forwarding burden.

CONFIDANT \cite{BB01} also detects misleading nodes by means of observation and more
aggressively informs other nodes of this misbehavior through reports sent around
the network.
Each node in the network hosts a {\em monitor} for observations,
{\em reputation records}
for first-hand and trusted second-hand reports,
{\em trust records} to control the trust assigned to received warnings, and
a {\em path manager} used by nodes to adapt their behavior according
to reputation information.  In more recent work \cite{BB02} \cite{BB03}, these researchers
find that reputation schemes can be beneficial for fast misbehavior detection,
but only when one can deal with false accusations,
for which they propose a solution using Bayesian statistics.  Our goal is to avoid
the machinery for managing these reports and their associated trust issues entirely.

Peer-to-Peer (P2P) networks face a similar situation in which they rely
on cooperation among self interested users. Recent studies have modelled
and quantified the incentives and disincentives for cooperation in P2P
networks (\cite{MF03},\cite{KL03}). These results generally appear to support
the feasibility of the approach.

Researchers have also investigated means of discouraging selfish routing
behavior in ad hoc networks, 
generally through payment schemes \cite{BH00},\cite{BH01},\cite{BHcellular}.
These approaches either require the use of tamper-proof hardware
modules or central bankers to do the accounting securely, both of which may not
be appropriate in some truly ad hoc network scenarios.
In the per-hop payment scheme proposed by Buttyan and Hubaux \cite{BH01},
the payment units are called {\em nuglets} and
reside in a secure tamper-proof module in each node.
They find that given such a module, increased cooperation is beneficial not only
for the entire network but also for individual nodes. 
We rely on much of
their work and likewise use a payment scheme.   In our simple ``chipcount''
mechanism, further described in Section~\ref{sec:selfish}, each node keeps track
of the number of packets it has
forwarded for its direct neighbors and expects corresponding willingness
from those neighbors to carry its traffic.  The scheme 
can result in unfairness to some hosts, but its simplicity and performance
may be appropriate in some scenarios.

\section{Overview of OCEAN}
\label{sec:overview}

OCEAN is a layer that resides between the network and MAC layers of the protocol
stack, and it
helps nodes make intelligent routing and forwarding decisions.
We have designed OCEAN on top of the Dynamic Source Routing Protocol (DSR) \cite{DSR},
although many of its principles may also be useful in other ad hoc routing
protocols.  In this section we describe the components of OCEAN that detect
and mitigate misleading routing behavior.  Section~\ref{sec:selfish} describes our techniques
for mitigating selfish behavior.

The OCEAN layer, which may reside on each node in the network, hosts five components:

{\bf NeighborWatch}: This module observes the behavior of the neighbors
of a node.  It relies on the omni-directional nature of the antenna
and assumes symmetric bi-directional links.  In particular, it tracks misleading
routing
misbehavior.  When forwarding a packet, the module buffers the packet checksum,
and then monitors
the wireless channel after sending the packet to its neighbor.  If it does not
hear the neighbor attempt to
forward the packet
within a timeout (default 1ms), NeighborWatch registers a negative event
against the neighbor node and removes the checksum from its buffer. On
the other hand, on overhearing
a forwarding attempt by the neighbor, NeighborWatch compares the packet to the
buffered checksum, and if it matches, it registers a positive event and removes
the checksum from its buffer.  If the checksum does not match, it treats the
packet as not having been forwarded.
These events are communicated to the RouteRanker, which maintains
ratings of the neighbor nodes.

The NeighborWatch module is not a guaranteed service.  It suffers from all the
same potential errors as the Watchdog \cite{MGLB00}, including, for example, the fact that
observing a neighbor forwarding a packet does not
guarantee that the packet is successfully received by the next node in
the route.

NeighborWatch on a node tracks only this one type of behavior
with neighbors directly interacting with it.
There are many other events among neighbors that NeighborWatch could potentially track,
but they are subject to too many vulnerabilities and thus
become more complex to analyze.
For instance, NeighborWatch on a node A could observe the success and
failure rates of its neighbors attempting to forward traffic between themselves.
Failure of a neighbor, B, to forward a packet from some other neighbor, C, could trigger
a negative event registration against B on A.  Unfortunately, only B knows for sure
whether its refusal to forward C's traffic is due to misleading behavior on B's part or is
instead B's legitimate response to C's previous misbehavior toward B.

{\bf RouteRanker}:
Every node maintains ratings for each of its neighboring nodes.  The rating is
initialized to {\em Neutral} and is incremented and decremented on
receiving positive and negative events respectively from the
NeighborWatch component.  We have found that the system performs more
satisfactorily when the absolute value of the negative decrement is larger than
the positive increment.  Once the rating of a node falls below
a certain threshold, {\em Faulty Threshold}, the node is added to
a {\em faulty list}. The faulty list represents a list of all
observed misbehaving nodes. A route is rated good or bad, based
on whether the next hop in the route belongs to the faulty list
or not. One can imagine a finer ranking between routes, where
good routes are further differentiated, but our simple
binary discrimination between good and bad
routes proves to be reasonably effective.
The default parameters we
use in OCEAN are tabulated in Table~\ref{tab:params}.
\begin{table}[t]
\begin{center}
\begin{tabular}{|c|c|}
\hline
Neutral Rating & 0 \\
Positive Step & +1 \\
Negative Step & -2 \\
Faulty Threshold & -40 \\
\hline
\end{tabular}
\end{center}
\caption{Default OCEAN Parameters}
\label{tab:params}
\end{table}

{\bf Rank-Based Routing}: The Rank-Based Routing module applies the
information from NeighborWatch in the actual selection of routes.
To make it possible to avoid routes containing nodes in the faulty list,
we add a variable-length
field to the DSR Route-Request Packet (RREQ) called the {\em avoid-list}.
The {\em avoid list} is a list of nodes that the RREQ transmitter
wishes to avoid in its future routes.
On re-broadcasting an RREQ, a node appends its faulty list to the
avoid list of the RREQ packet.
Any node receiving an RREQ checks the RREQ avoid list.
Depending upon the avoid list and the RREQ-route, a node decides
whether to suppress the RREQ, or honor the RREQ by either re-broadcasting it
or replying with a DSR Route-Reply.
If the intersection of the avoid list and the DSR route in the RREQ
packet is non-void (i.e. a node which was requested to be avoided is
in the route), the RREQ packet is suppressed. Similarly, a DSR
Route-Reply (RREP) is honored only if the route in the RREP does not
contain a node in the locally-maintained faulty list. Otherwise, the
RREP is simply dropped.

In this way, each node along the route makes
its own local decision about nodes to trust, and a node has control only over routes
that go directly through it.  Nodes may tamper with the
avoid lists, in particular with a {\em Rushing Attack},
but in Section~\ref{sec:simresults} we describe the attack and show
through simulation results that
the protocol is fairly robust against this kind of avoid list tampering.
The avoid list could also be made tamper-proof with increased overhead, in
the context of cryptographically secure protocols.

{\bf Malicious Traffic Rejection}:
This module performs the straight-forward function of rejecting
traffic from nodes it considers misleading. We employ
the policy of rejecting {\em all} traffic from a misleading
node so that a node is not able to relay its
own traffic under the guise of forwarding it on somebody else's behalf.

{\bf Second Chance Mechanism}:
The Second Chance Mechanism is intended to allow nodes previously considered misleading
to become useful again.
Without a second chance, once a node is added to the faulty list, all future
routes through it would be avoided, giving that node no opportunity
to demonstrate its ``goodness.''  This may be unfairly harsh
on a node, especially since the NeighborWatch module is not guaranteed to
be correct in its judgments.  A node may simply have been experiencing
transient link failures,
or it may have needed to restart its network interface. To account for
such problems, we use a timeout-based approach where a misleading node is removed
from the faulty list after a fixed period of observed inactivity.
Even though
the node is removed from the faulty list, its rating
is not increased to neutral, so that it can quickly be added back
in the event of continued misbehavior. This timeout value is called the
{\em Faulty Timeout}.

\section{Selfish Behavior}
\label{sec:selfish}
In this section we describe how OCEAN attempts to mitigate selfish
routing behavior in ad hoc networks.  The general idea is to punish
nodes for their selfish behavior, by rejecting their traffic, in the
hopes that this threat will act as a deterrent.

Unfortunately, it can be hard to detect selfish behavior through
observations of the routing protocol itself, since there are
many undetectable techniques through
which a node can avoid becoming visible as a potential router.
A few examples of such manipulation of routing-related messages
(specific to DSR), which
can go undetected, are
\begin{enumerate}
\item Dropping the Route Request (RREQ). 
Inappropriate dropping of  RREQs can easily go undetected, since RREQs are broadcast,
are not acknowledged, and can legitimately be dropped in some circumstances.
\item Adding too many nodes to the route in the RREQ packet.  If a node adds many other
nodes to a RREQ packet, the resulting path will appear to be undesirably longer
than paths through nodes that have not tampered with the RREQ.  The result is that
paths through such a nodes are unlikely to be chosen.
\item Adding a non-existent route in the RREQ packet.  If a node modifies the
path in the RREQ such that it does not exist, then traffic will not
reach the misbehaving node.
\end{enumerate}

Because such manipulations of the routing protocol may go undetected, if
not otherwise secured through heavier-weight cryptographic means, the best
evidence of a node's cooperation is
the actual number of packets it forwards.  A decision about whether
to forward a node's packet can be based on its past forwarding performance.
This creates a loose packet-forwarding economy between nodes.
Similar economy-based approaches \cite{BH00},\cite{BH01},
\cite{BHcellular}
require the use of tamper-proof hardware modules or centralized bankers for
secure accounting.

OCEAN instead relies only on direct observations of interactions
with neighbors to measure their performance.
Every node maintains counters
called {\em chipcounts} for each neighbor. A chipcount can be thought
of as a bank balance in a bank hosted by the node that maintains the
chipcount. A node {\em earns} chips at a node upon forwarding a packet
on behalf of that node.  Similarly, a node {\em loses}
chips with a node it asks the node to forward a
packet.  When deciding whether to service a forwarding request, a
node checks its chipcount for that neighbor.  If the chipcount
falls below a threshold, the node denies the request.

We have experimented with two trade-based schemes:
{\em optimistic} and {\em pessimistic}.
In the optimistic scheme, a node A increments the chipcount
for a node B whenever node B accepts a packet from node A,
regardless of whether B actually forwards the packet.
The pessimistic scheme, on the other hand, increments the chipcount
for node B only when node B is observed to forward the packet.
In both cases, node A will only know to ask B to forward
traffic if B has previously participated in the route
request protocol and is therefore on a route through A.  If B fails to
forward A's traffic at this point, then the NeighborWatch module will
detect the misleading behavior.  The chipcount scheme, instead,
detects behavior where B selfishly asks A to forward traffic, even
if B has managed to escape being on any routes through A.

The pessimistic scheme suffers from a deadlock problem where two
nodes may not forward packets for each other for a long time because
one node fails to observe the other node forwarding a packet on
his request. On the other hand, an optimistic scheme could be too lenient
on the misbehaving nodes, since the chips for these nodes are
incremented even though the nodes do not forward packets.

This trade-based scheme may suffer from unfairness to nodes on the periphery
of the network.  These nodes may be punished because they do not receive sufficient
opportunity to forward packets for others. This can cause the throughput of the network
to fall significantly, and it can cause the network to ``shrink'' in upon itself
as the interior neighbors of the peripheral nodes themselves appear to become
peripheral nodes.
To address this problem, we add a tunable parameter
called the {\em Chip Accumulation Rate} (CAR). CAR is the rate at
which all chipcounts in the network are increased per unit time. Thus,
even when a neighbor does not forward any packets for a node, it will eventually
have a non-zero chipcount value and can thus forward traffic at some reduced
rate.

Setting an appropriate CAR value requires a trade-off.
An infinite CAR value would allow selfish nodes to enjoy full
freedom in relaying their packets, since they will never run out of
chips at any node.  On the other hand, a zero
CAR value suffers from the unfairness to peripheral nodes and the reduced
network throughput for cooperating nodes described above.  In our simulations, using
high mobilities and a topology that places many nodes on the periphery,
we find that having a low CAR value punishes
selfish nodes much more than the cooperating nodes, which indicates
that at least the trade-off plays in our favor.

\section{Simulation Results}
\label{sec:simresults}
\begin{table}[t]
\begin{center}
\begin{tabular}{|c|c|}
\hline
Number of Nodes & 40 \\
Maximum Velocity & 20 m/s \\
Dimensions of Space & 1500m x 300m \\
Nominal Radio Range & 250m \\
Connection Life & 8 packets\\
Min Connection Length & 2 hops\\
Source Data pattern (each) & 4 packets/second \\
Application Data Payload Size & 64 bytes/packet \\
Raw Physical Link Bandwidth & 2Mbps \\ \hline
\end{tabular}
\caption{Parameters for OCEAN Simulations}
\label{tab:simparam}
\end{center}
\end{table}
In this section, we demonstrate that mitigating routing misbehavior and
selfish behavior through the use of direct observation of nodes can
often be as effective as techniques deploying second-hand node reputation.
We first describe our simulation
environment and then go on to compare the throughput of OCEAN
in the presence of misleading or selfish nodes against a network using
second-hand reputation information.
\begin{figure}[htb]
  \centering
   \centerline{\epsfig{figure=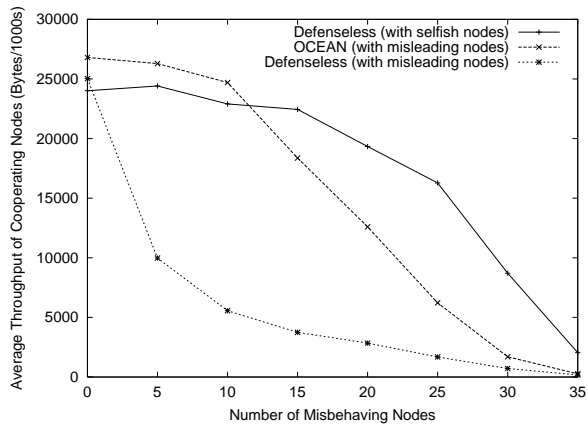,width=8cm}}
\caption{Average Throughput of the Cooperating Nodes for varying numbers
of misbehaving nodes in the Network. OCEAN can sustain a good throughput
for high percentages of misbehaving nodes.}
\label{fig:thput_vs_num_mal}
\end{figure}

We implemented OCEAN and its variants in {\em GloMoSim} \cite{GL98},
a commonly-used simulator in the ad hoc research community.
These simulations model radio propagation using the realistic
{\em two-ray ground reflection} model and account for physical
phenomena such as signal strength, propagation delay, capture effect
and interference. The Medium Access Control protocol used is the
IEEE 802.11 Distributed Coordination Function(DCF). The parameters
we use for the simulations are given in Table \ref{tab:simparam}.
All results have been
plotted after taking an average over 20 simulation runs. The ratio of the
standard deviation and the average was around 1.35 for highly mobile
scenarios and around 0.70 for low mobility scenarios.
We simulate only connections
longer than two hops, since one-hop connections would artificially inflate
our throughput figures (even in the face of 100\% misleading nodes, we would
see some throughput.)
While the mobility model we simulate is not as realistic as we could hope for,
it is very commonly used, making it possible for us to compare our work
directly against others' results.  We hope to test our ideas in other models
and real systems in the future.

\begin{figure*}[t]
\begin{center}
\subfigure{\epsfig{figure=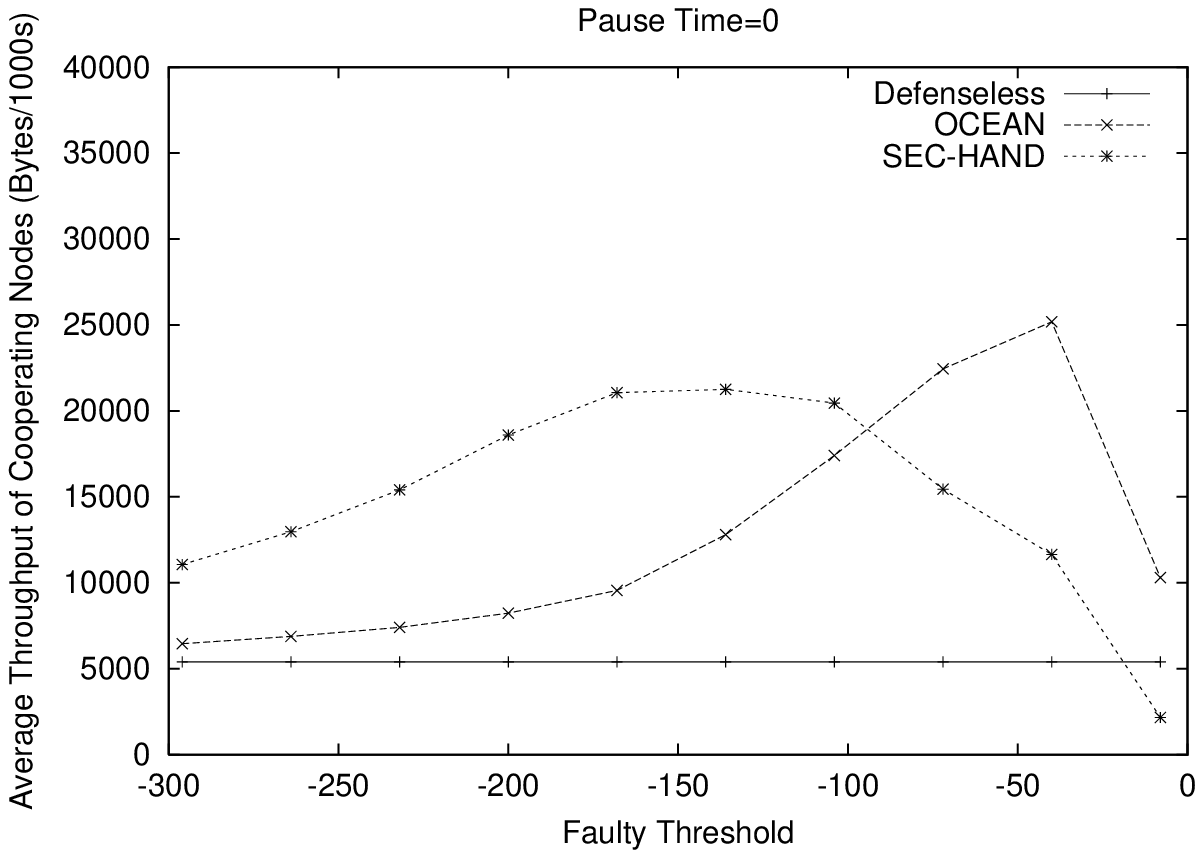,width=8cm}}
\subfigure{\epsfig{figure=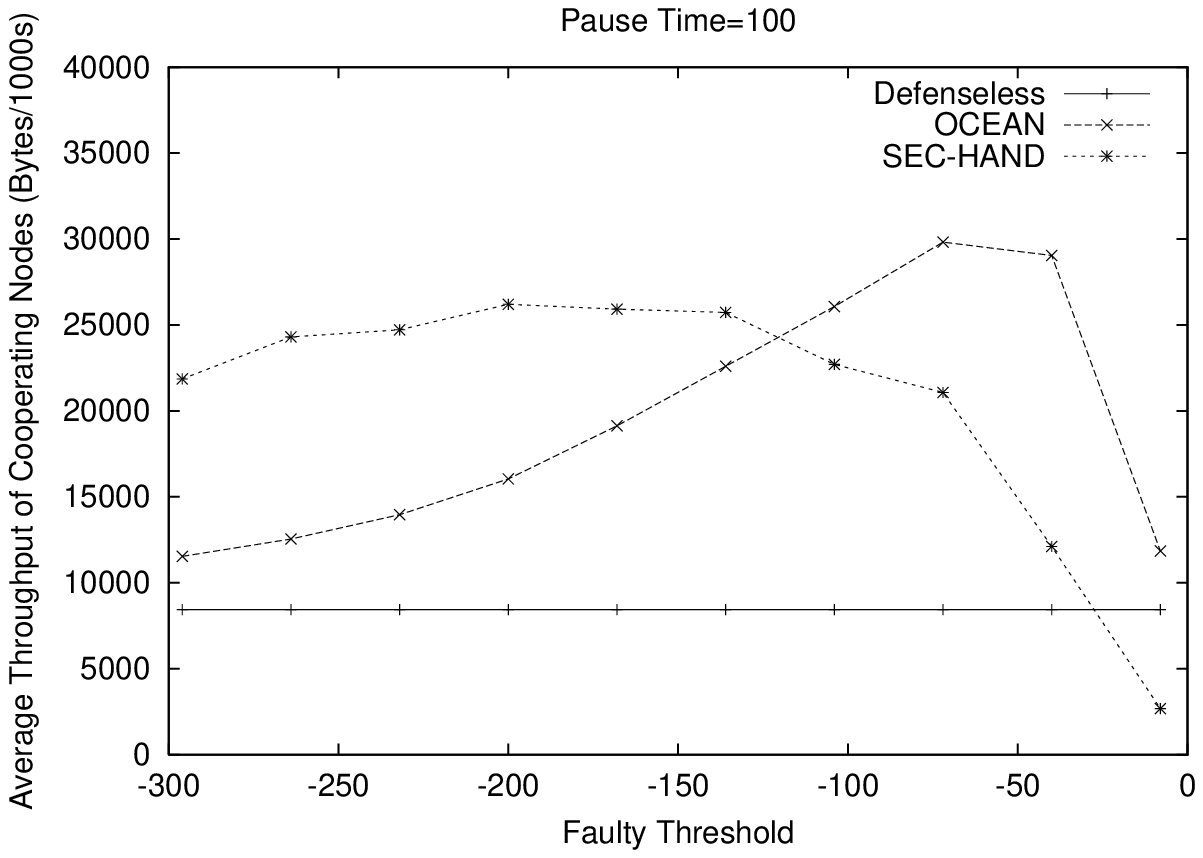,width=8cm}}\\
\subfigure{\epsfig{figure=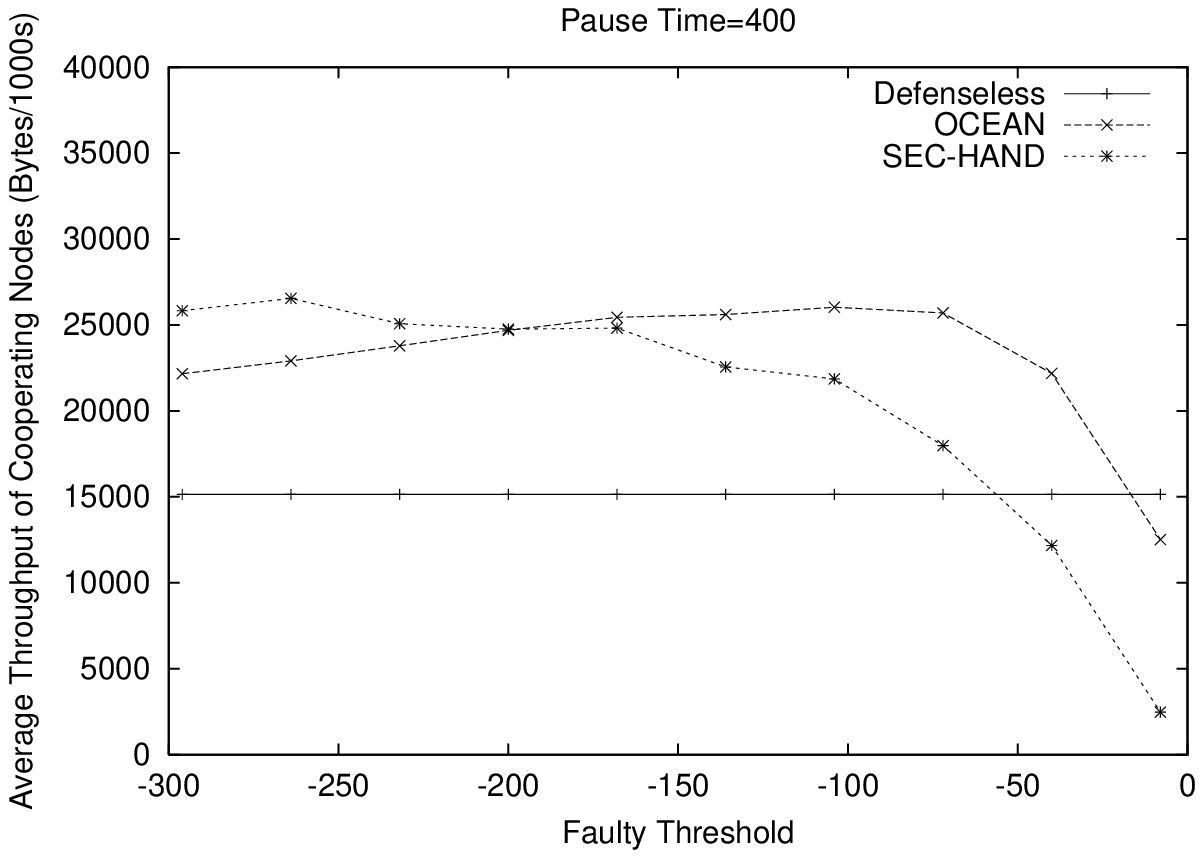,width=8cm}}
\subfigure{\epsfig{figure=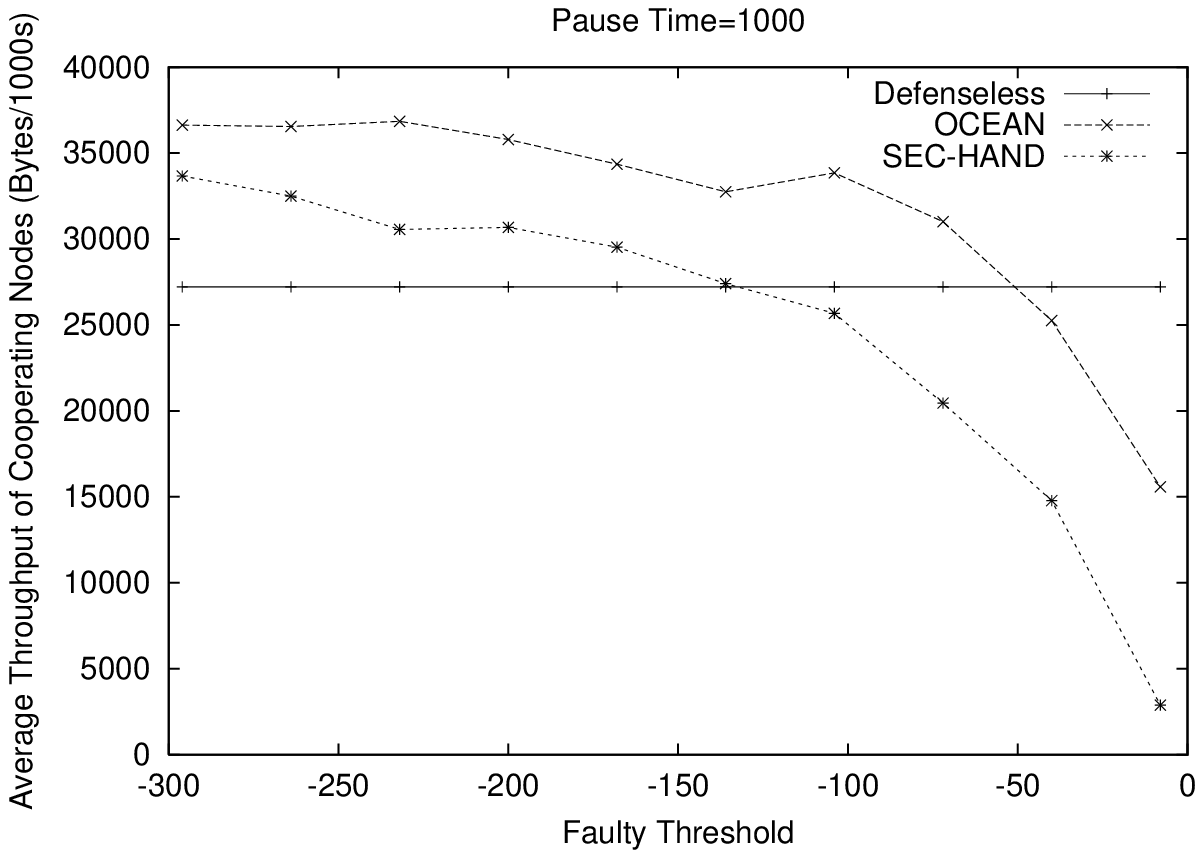,width=8cm}}
\caption{Comparison of Direct Observation (OCEAN) and
Using Second-Hand Information. The number of misleading nodes is 10.  The lower the
Pause Time, the more mobile the network, with a zero Pause Time implying
implies continuous mobility.}
\label{fig:faulty_thresh}
\end{center}
\end{figure*}

We first consider the throughput of cooperating nodes in OCEAN in the presence of
varying numbers of misleading nodes and compare it to the same network without
OCEAN and also the same network without OCEAN
but in which the nodes are merely selfish rather than misleading
(Figure \ref{fig:thput_vs_num_mal}).  We make this latter comparison because it helps
us judge the potential utility of OCEAN.  Any misbehavior-detecting protocol
should not be expected to perform better than a network in which the nodes drop
packets but do not actively mislead other nodes.  This is because misbehavior detection can
at most prevent nodes from being misled by others; it cannot force the misbehaving
nodes to begin forwarding packets.
We observe that OCEAN performs drastically better than the same network
without OCEAN, and it can even sustain a performance close
to the defenseless network with merely selfish nodes.  It helps
sustain 90\% of the original throughput even when 25\% of the nodes
misbehave.  As the percentage of the misbehaving nodes approaches
100\%, the throughput inevitably falls to zero.  At lower numbers of misbehaving
nodes, OCEAN actually appears to perform better than the network with
merely selfish nodes.  This is because OCEAN also routes around nodes that drop
packets because they are merely overloaded, not intentionally misleading.

We next compare the performance of OCEAN to a protocol, SEC-HAND, that
uses second-hand reputation information.
SEC-HAND is intended to represent the family of
protocols that use second-hand reputation information.
SEC-HAND uses ALARM messages between nodes to communicate
information about misbehaving nodes (similar to those used in CONFIDANT \cite{BB01}).
We augmented the DSR Route-Error Packet
to contain an ``Alarm'' field. The node in the Alarm
field is advertised as misbehaving and all nodes
overhearing the Alarm add the accused node
to their respective faulty lists.  SEC-HAND is otherwise
like OCEAN, to make it possible to compare the
techniques fairly.

We make our comparisons of OCEAN and SEC-HAND across varying
values of the Faulty Threshold and
varying degrees of mobility (in Figure \ref{fig:faulty_thresh}).
We vary the Faulty Threshold, because it controls
the speed and the accuracy
of misbehavior detection.  A small (by absolute value) faulty threshold adds
nodes faster to the faulty list, but also suffers from the problem of
false positives.  A large faulty threshold suffers from a slow detection
speed.  Detection speed is particularly important for OCEAN,
since it needs to evaluate new neighbor nodes from scratch.  Hence,
faster detection should help OCEAN.  SEC-HAND should tolerate slower
detection speeds, since it keeps records of remote nodes and thus has
more information available when a new node joins the neighborhood.
Accuracy, on the other hand
is critical for SEC-HAND, since SEC-HAND can spread
false information in the network if the detection was not accurate.
OCEAN should be more resilient to false detection, since bad
information will be kept local.

We vary the degree of mobility, because
we would expect SEC-HAND to perform better than OCEAN in
highly mobile scenarios.
High mobility implies a quickly changing neighborhood,
and OCEAN nodes must
must learn about new neighbors from scratch.  On the other hand, SEC-HAND
maintains ratings for remote nodes, which can be helpful in quickly
judging new local nodes that were previously remote.

From the results, we find that at high Faulty Thresholds,
SEC-HAND is indeed able to perform
better than OCEAN at high mobilities, as expected.
On the other hand, OCEAN outperforms
SEC-HAND at low Faulty Thresholds because SEC-HAND
is much more susceptible to false positives.
At high mobility, OCEAN is more sensitive to the
tuning of the Faulty Threshold parameter, while SEC-HAND
performs well over a broader range of tunings.
Both protocols perform better with lower mobility.
Overall, if OCEAN and SEC-HAND both tune the faulty thresholds
to suit themselves, OCEAN can outperform SEC-HAND (even for highly
mobile networks).
We conclude that, even in highly mobile networks, the network can perform
reasonably well without the need to exchange second-hand information.

\begin{figure*}[t]
\begin{center}
\subfigure{\epsfig{figure=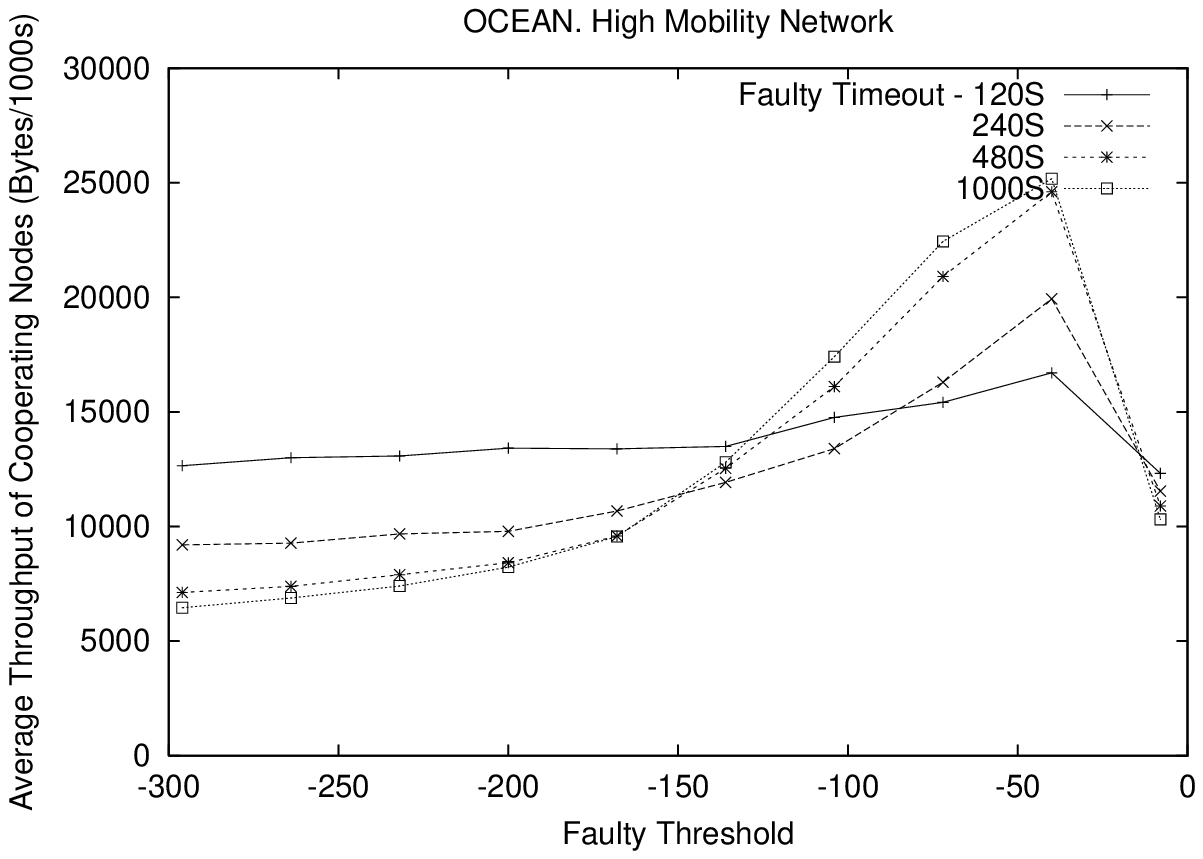,width=8cm}}
\subfigure{\epsfig{figure=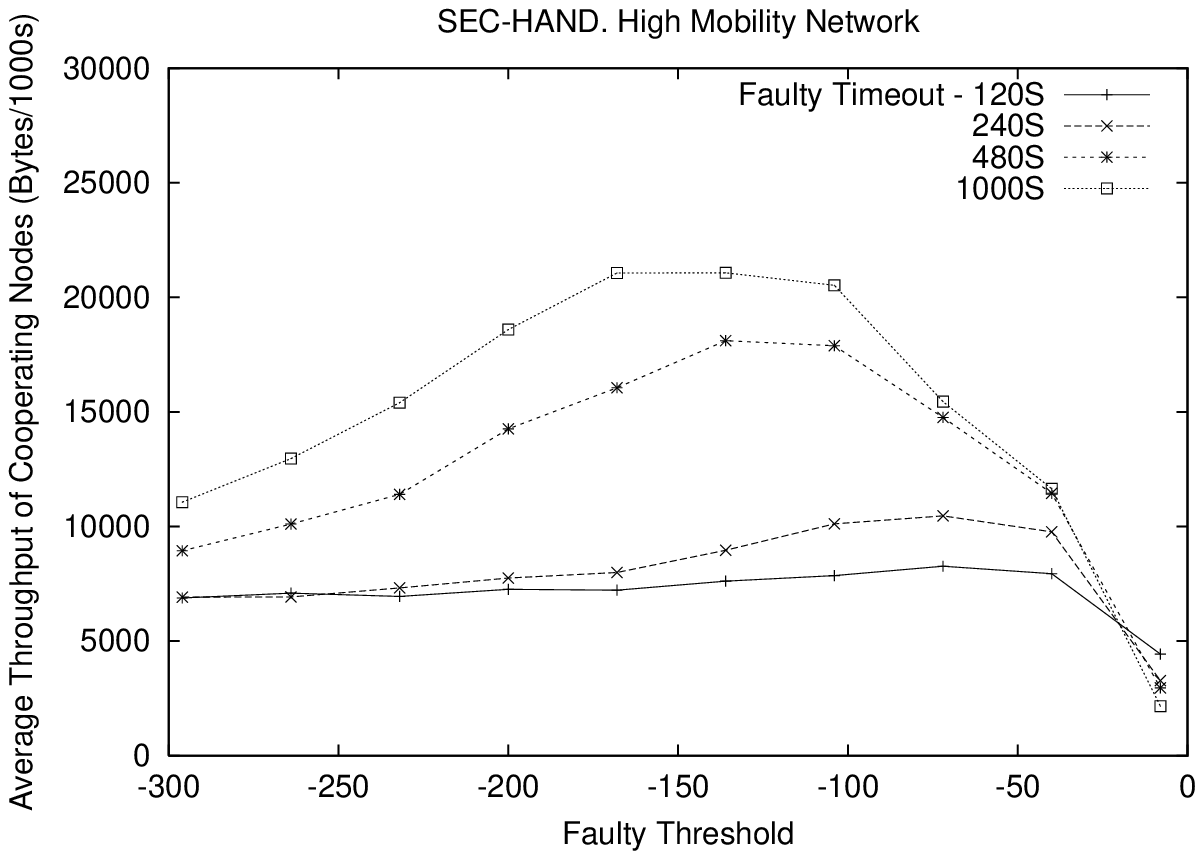,width=8cm}}
\caption{Average Throughput of Cooperating Nodes with varying Faulty Timeout
and Faulty Threshold parameters in High Mobility Scenarios (Pause Time=0).
OCEAN is more resilient to lost information due to Faulty Timeouts.}
\label{fig:thput_vs_timestamp_expire}
\end{center}
\end{figure*}

We further compare the performance of OCEAN and SEC-HAND
in the face of transient failures (weak links) in Figure \ref{fig:thput_vs_timestamp_expire}.
Both of the protocols can incorrectly detect such failures as misbehaving node behavior
and over-react accordingly.  Because of this problem, OCEAN includes a parameter
called Faulty Timeout, which controls the
{\em idle time} before a neighbor declared misbehaving is given a second
chance and is elevated to the status of being non-faulty (albeit with a low rating).
Some protocols using second-hand reputation information never give nodes
a second chance, but we also implement this feature in 
SEC-HAND, to provide fair comparison.  A concern, however, is how
quickly the protocols respond when the node given a second chance still
proves to be misbehaving.  In SEC-HAND,
if the timed-out faulty node is multiple hops away,
we need to wait for another ALARM message before the
node is added back to the faulty list. In
OCEAN, we only detect misbehavior of direct neighbors, which allows
us to determine quickly, on subsequent traffic through them, whether to
put misbehaving neighbors back on the faulty list.
This is seen in a comparative degradation in SEC-HAND's
performance at low Faulty Timeout values.

\begin{figure*}[htb]
\begin{center}
\subfigure{\epsfig{figure=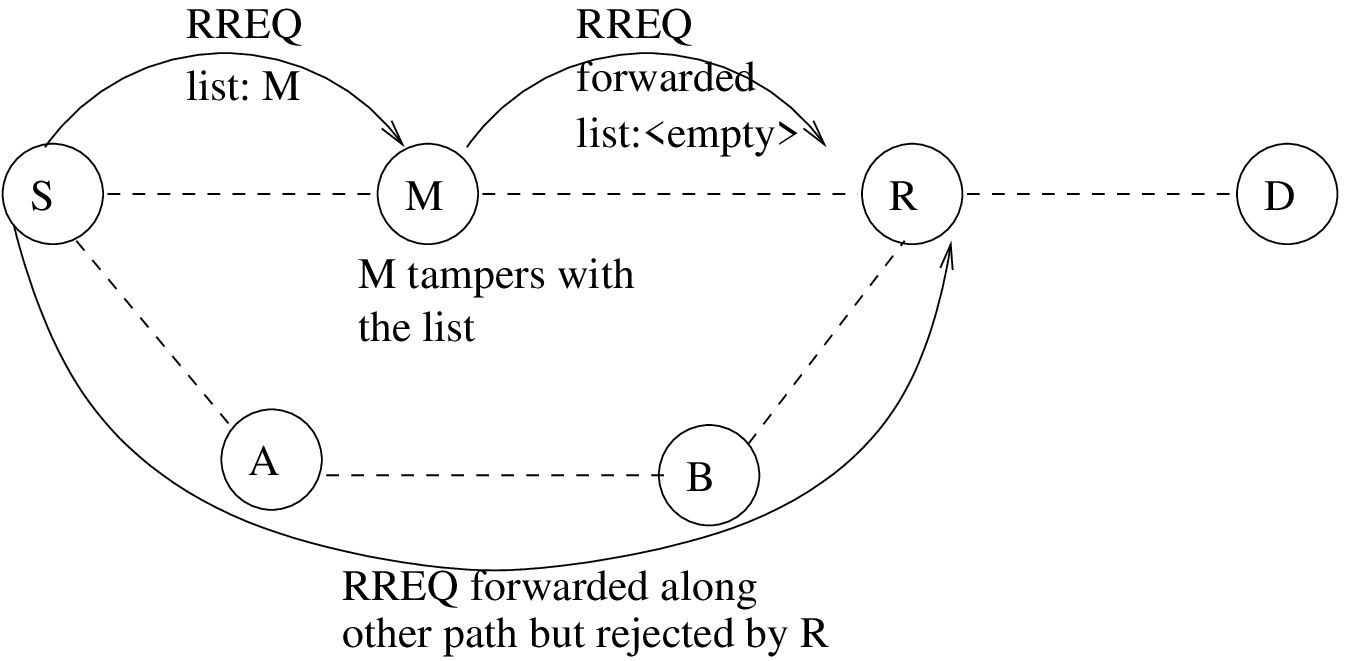,width=8cm}}
\subfigure{\epsfig{figure=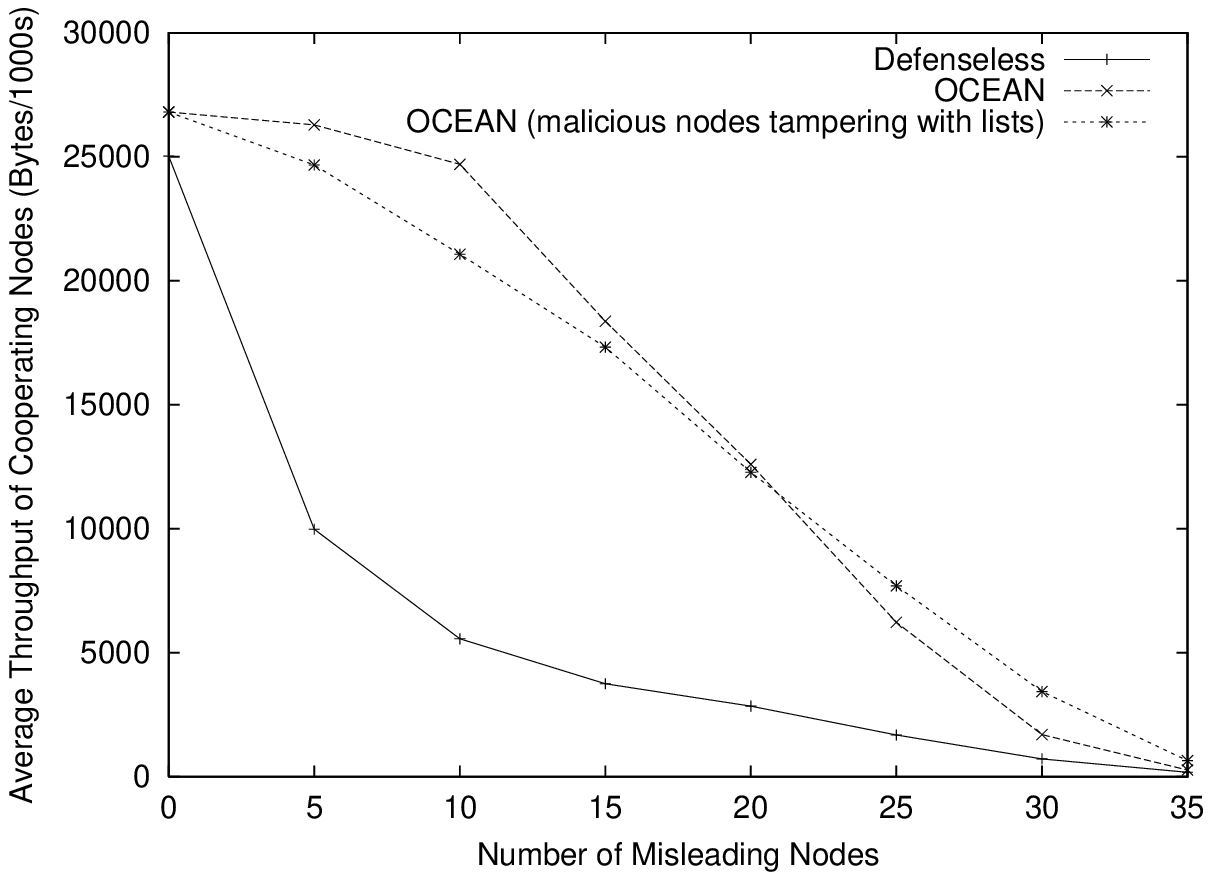,width=8cm}}
\caption{The figure shows how a rushing attack could be mounted by
manipulating avoid lists.  However, the attack makes little difference
in random ad hoc networks}
\label{fig:rushing}
\end{center}
\end{figure*}

We next examine the vulnerability of OCEAN if it is deployed over a protocol
that does not secure control packets against tampering.
The attack that
a malicious node may attempt is called a {\em Rushing Attack} \cite{HPJ01}, whereby
a node hurries a ``tampered'' route-request through itself.  The next node
along the path will forward this tampered route request and drop further
instances of the same route request that come from other nodes.
In this way, the malicious node
can establish a route through itself because its route is the first
seen by a downstream node, and it can later drop the data packets
sent through it.  The {\em Rushing attack}
in OCEAN is illustrated in Figure \ref{fig:rushing}.  Note, though that if node R
in the illustration had been the destination, the problem would have been
avoided, since by default DSR requires destination nodes to reply to all
route requests they receive.  In that case, the {\em good route} would have been found.
However, in our simulations of random networks, this attack seemed
to make little difference.  Figure~\ref{fig:rushing}
also shows the throughput of the network
when the malicious nodes tamper with the
avoid lists.  The throughput remains reasonable, since
for the attack to be possible a relatively specific
configuration of nodes is needed, which does not occur
frequently, at least in random ad hoc networks.
There are many other ways a malicious node
may attack a network if the routing layer is not secured. However, we believe our
experiments show that OCEAN does not add any {\em new} vulnerabilities that should
significantly affect performance.

\begin{figure}[htb]
  \centering
   \centerline{\epsfig{figure=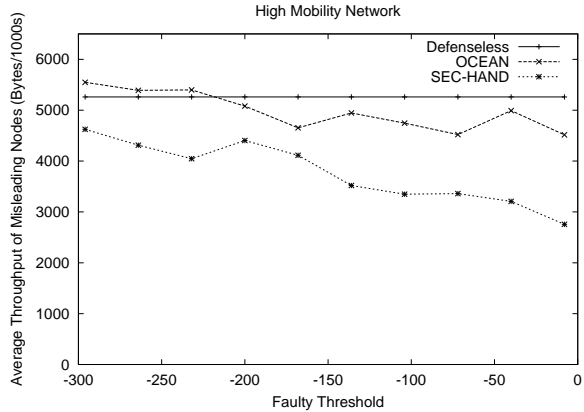,width=8cm}}
\caption{Average Throughput of Misleading Nodes with varying Faulty Threshold.
The number of misleading nodes is 5. Since the misleading nodes also
use OCEAN, they intelligently route around nodes that do not forward their
traffic.}
\label{fig:malthput_vs_num_mal}
\end{figure}

Another metric of evaluation for OCEAN is the throughput of the misleading nodes.
Ideally, we would like the throughput of the misleading nodes to be as
low as possible, to deter their behavior. Figure \ref{fig:malthput_vs_num_mal} plots the
throughput of the misleading nodes in defenseless, OCEAN, and SEC-HAND networks.
Unfortunately, we see that OCEAN is not very effective in thwarting the throughput
of the
misleading nodes.  This is because the misleading nodes also use OCEAN to
route around other misleading nodes or nodes
that did not forward their packets.
They were thus able to maintain a good (and sometimes better) throughput even
when the network was using OCEAN. Even in
SEC-HAND the misleading nodes are able to take advantage
of the SEC-HAND modules, however SEC-HAND is better at punishing
the misbehaving nodes,
since the bad reputation of the misbehaving nodes spreads much faster. 

\begin{figure*}[htb]
\begin{center}
\subfigure{\epsfig{figure=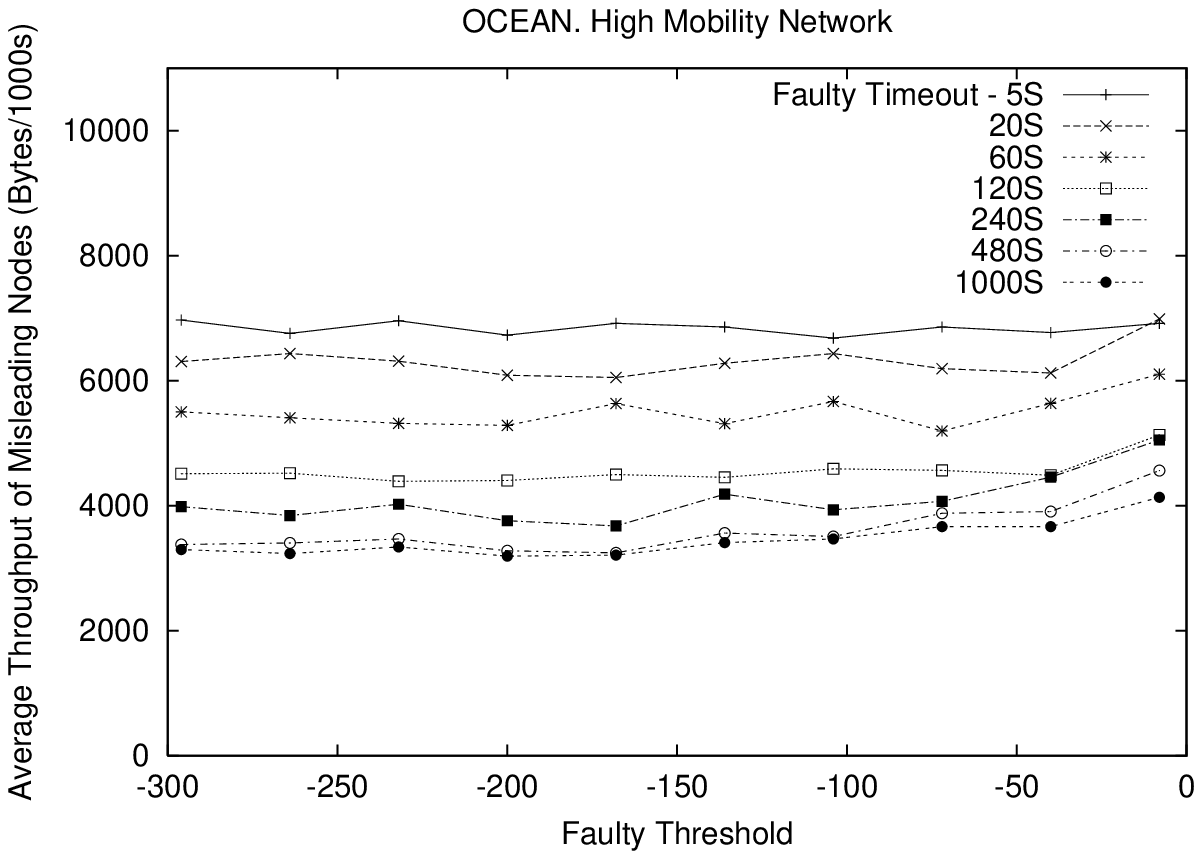,width=8cm}}
\subfigure{\epsfig{figure=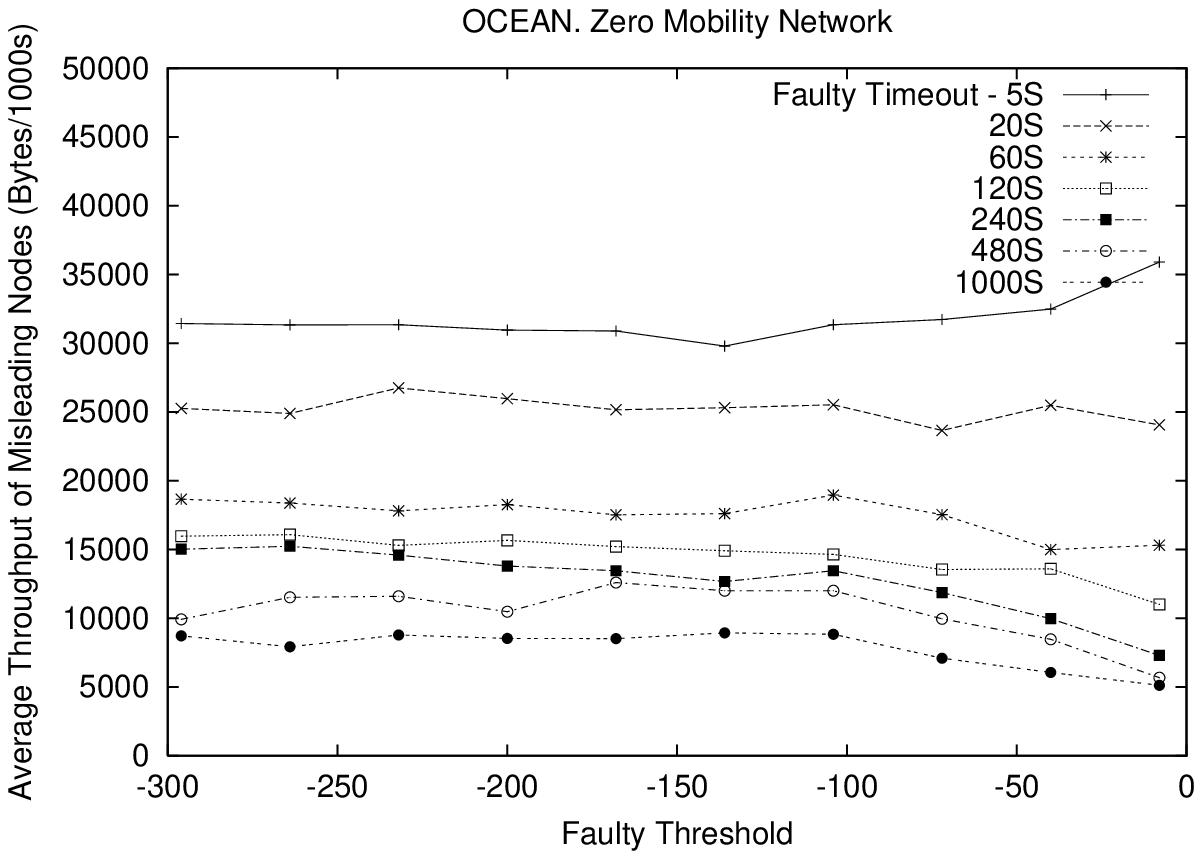,width=8cm}}
\caption{Average Throughput of Misleading Nodes with varying Faulty Timeout
and Faulty Thresh parameters.  The throughput of the
misleading nodes increases with a decreasing Faulty Timeout, since a lower
Faulty Timeout gives the misleading nodes more chances.  Note the drastically
different scales on the vertical axes of the two plots.}
\label{fig:malthput_vs_timestamp_expire}
\end{center}
\end{figure*}

We go on to examine the sensitivity of the throughput of these misleading nodes
to varying Faulty Threshold and Faulty Timeout values.
The results are in
Figure~\ref{fig:malthput_vs_timestamp_expire}. Interestingly, the
throughput of the misleading nodes is almost constant with a varying Faulty
Threshold. Since the malicious nodes also use OCEAN with the same
parameters as the rest of the network, the positive and
negative effects of decreasing the Faulty Threshold
almost cancel each other out. On the other hand, increasing the
Faulty Timeout thwarts the throughput of the faulty nodes, as one
would expect, since the misleading nodes get fewer second chances.

\begin{figure*}[htb]
\begin{center}
\subfigure{\epsfig{figure=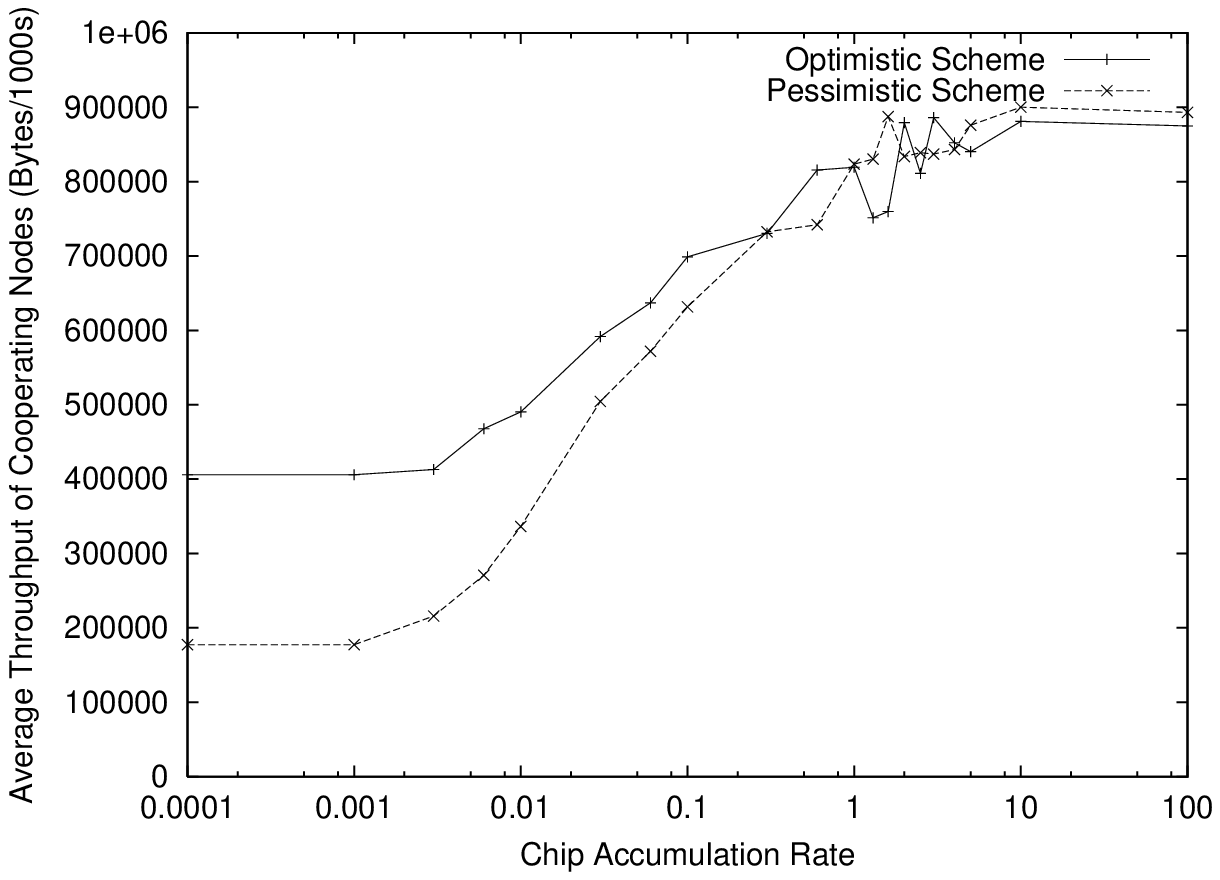,width=8cm}}
\subfigure{\epsfig{figure=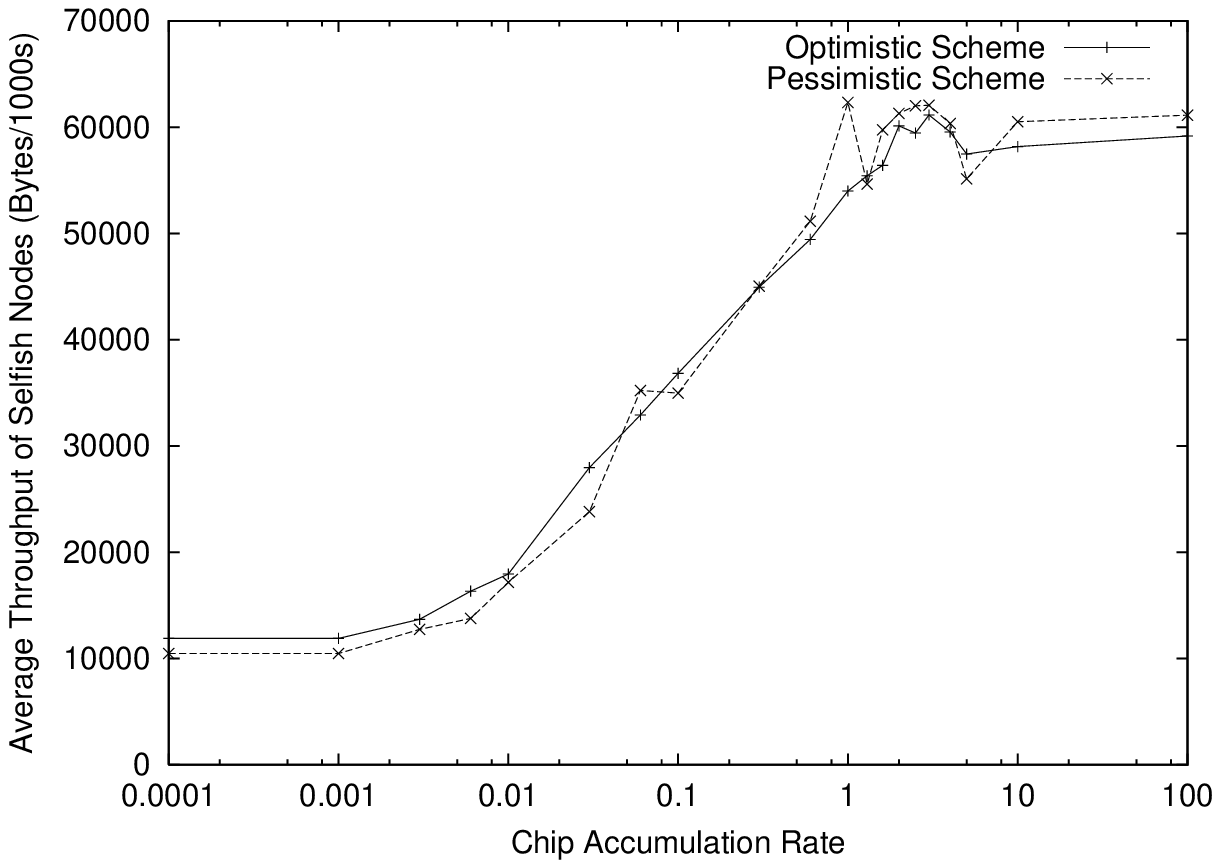,width=8cm}}
\caption{Throughput of the cooperating nodes and selfish nodes
in the network with
varying chip accumulation rates. Number of selfish nodes=5.  Note the drastically
different scales on the vertical axes of the two plots.}
\label{fig:thput_nuglets}
\end{center}
\end{figure*}

Finally, in addition to considering networks with misleading routing behavior,
we consider the performance of networks containing merely selfish nodes.
We study the performance of
the economy-based scheme proposed in Section~\ref{sec:selfish} in a simulation
that places many nodes on the periphery of the network (a 1500m by 300m rectangle).
This topology emphasizes the problems that our economy scheme causes in terms of
unfairness to peripheral nodes and reduced throughput for cooperating nodes.
Figure
\ref{fig:thput_nuglets} plots the throughput of the cooperating
nodes and the selfish nodes with varying chip accumulation
rates (CARs) under optimistic and pessimistic schemes.

At a low CAR
value, the throughput of the
cooperating nodes suffers
a two-fold decrease, leading one to want to tune CAR to higher values.
On the other hand, the throughput of selfish nodes
changes by a factor of five to six, leading one to want to
tune CAR to lower values, to adequately punish selfish nodes.
Overall, we see that an
optimistic scheme better suits these trade-offs than a pessimistic scheme.
An ``optimum'' CAR value, though, may vary depending on network
requirements, and there is no good mechanism in our simple scheme for
preventing individual nodes from tuning CAR to whatever value best
meets their selfish needs.

\section{Authentication Issues}
\label{sec:discussion}

In the work we present in this paper,
we assume that nodes do not spoof each
other's identities, since this would allow
misbehaving nodes to exploit the good reputation
of neighboring cooperating nodes.
Since in reality it is easy to spoof IP addresses
and even MAC addresses, this would imply the
use of a cryptographically secure authentication
mechanism, perhaps as provided through a secure
routing protocol.

Unfortunately, we do not yet find a secure routing protocol
that handles authentication in a manner that matches the
spirit of OCEAN in being truly ad hoc and also manageably
simple.  Some secure routing
protocols rely on pre-assigned certificates from common certificate
authorities to authenticate nodes \cite{Roy02}, but it may
not always be possible in truly ad hoc contexts for nodes
to hold such pre-assigned certificates from authorities that
all nodes will respect.  Efforts to develop on-the-fly
certificate authorities within the network \cite{HPJ01} \cite{ZH99} appear either
to be quite complex or to distinguish
the role of certificate authority among a subset of nodes, which does not provide
complete decentralization.

In OCEAN, we instead hope to give up the requirement for stable identities for nodes
\emph {at the routing level},
and instead merely prevent nodes from spoofing one another.
To achieve this objective, 
nodes generate their own asymmetric key pairs, the
public portion of which they can exchange with neighbors,
using them perhaps merely to agree upon lighter-weight authentication
secrets.

This mechanism, however, does not prevent a node from generating
multiple identities rapidly and then discarding identities once
the associated reputations have fallen below threshold.
To deal with such short-term identities, we hope to
leverage recent work on {\em proof-of-effort} mechanisms \cite{Abadi} \cite{Dwork}.
In the context of OCEAN, we could
mandate that a new identity be accepted only
if the identity-owner shows reasonable proof of recent effort in
generating that identity.
This would not require nodes to maintain stable identities for
very long, but they would not find it advantageous to cycle through
them fast enough to cause much havoc undetected.

\section{Conclusions and Future Work}
\label{sec:conclusion}

This paper presents the OCEAN techniques for detecting and
mitigating misleading routing behavior in ad hoc networks.
Our goal was to study how far we can
get using only direct observations of neighbors.  We find
that this scheme works surprisingly well, in terms of
network throughput, considering its
simplicity compared to schemes that share second-hand
reputation information throughout the network.  Compared
to such reputation schemes, OCEAN is
more sensitive to the tuning of some parameters, and it
fails to punish misbehaving nodes as severely, but
it performs almost as well, and sometimes even better, across a wide
range of degrees of mobility.

We also find that our {\em chipcount} scheme provides a
simple first step at being able to deter selfish behavior
in the network.
However, this scheme is
accompanied with network throughput deterioration.
Further work is warranted to see if we can do a better
job of using only
directly-observable information to identify and deter
selfish nodes without such significant reductions in the throughput
of cooperating nodes.

We also understand that our random simulation models cover an
unrealistically small sample of potential network behavior.
We would like to simulate other more realistic models, and
test our ideas in real systems if we have sufficient resources
to do so.

Finally, we plan to study how we can provide
more effective infrastructure-free authentication
in ad hoc networks assuming that identities
need not be entirely stable at the routing level,
but that spoofing of other nodes is unacceptable.

\section{Acknowledgments}
We thank the students in CS444N for their initial efforts and
input into this project.  We especially
thank Yan Liu who continued to work on this project after the class was over.
We are very grateful to the Nokia Research Center for supporting this project.
The project has also been supported in part by MURI award number F49620-00-1-0330
and the Stanford Networking Research Center.

\nocite{*}
\bibliographystyle{IEEE}

\begin{thebibliography}{1}
\small{
\bibitem{Abadi} M. Abadi, M. Burrow, M. Manasse and T. Wobber.  Moderately Hard,
Memory-bound Functions.  In {\em Proceedings of the 10th Annual network and Distributed
System Security Symposium (NDSS)}, February 2003.

\bibitem{HashCash2} A. Back. HashCash. Available on the web at URL
{\texttt www.cypherspace.org/~adam/hashcash}

\bibitem{PGPOnWireless} M. Brown, D. Cheung, D. Hankerson, J.L.
Hernandez, M. Kirkup and A. Menezes. PGP in Constrained
Wireless Devices. In {\em 9th USENIX Security Symposium}, Aug. 2000.

\bibitem{BB01} S. Buchegger and Jean-Yves Le Boudec. Performance
Analysis of the CONFIDANT Protocol; Cooperation of Nodes - Fairness in Dynamic
Ad Hoc NeTworks. In {\em Proceedings of IEEE/ACM Symposium on Mobile Ad Hoc
Networking and Computing (MobiHOC)}, Lausanne, CH, June 2002.

\bibitem{BB02} S. Buchegger and Jean-Yves Le Boudec. The Effect of Rumor Spreading in Reputation Systems for Mobile Ad Hoc Networks.  In {\em WiOpt'03: Modeling and Optimization
in Mobile, Ad Hoc and Wireless Networks}, March, 2003.

\bibitem{BB03} S. Buchegger and Jean-Yves Le Boudec. Coping with False Accusations in Misbehavior Reputation Systems for Mobile Ad Hoc Networks.  EPFL Technical Report Number IC/2003/31, 2003.

\bibitem{BH00} L. Buttyan and J.P. Hubaux. Enforcing Service Availability
in Mobile Ad hoc WANs. In {\em Proceedings of IEEE/ACM Workshop on Mobile Ad Hoc
Networking and Computing (MobiHOC)}, Lausanne, CH, June 2002.

\bibitem{BH01} L. Buttyan and J. Hubaux. Stimulating Co-Operation in Self
Organizing Mobile Ad Hoc Networks. Technical Report DSC/2001/046, EPFL-DI-ICA,
August 2002.

\bibitem{HashCash3} camram. Available on the web at URL
www.camram.org, 2002.

\bibitem{HashCash1} C. Dwork and M. Naor. Pricing via processing or
combating junk mail. In {\em Advances in Cryptology--CRYPTO '92}, pages
139-147. Springer, 1999.

\bibitem{Dwork} C. Dwork, A. Goldberg, and M. Naor.  On Memory-Bound Functions for
Fighting Spam. Manuscript, 2002.

\bibitem{MF03} M. Feldman, K. Lai, J. Chuang, I. Stoica. Quantifying
Disincentives in Peer-to-Peer Networks. In {\em Workshop on Economics
of Peer-to-Peer Systems}, Berkeley, 2003.

\bibitem{KeyManagement} K. Fokine. Key Management in Ad Hoc Networks. Thesis.

\bibitem{HJP02} Y. Hu, A. Perrig and D.B. Johnson, SEAD: Secure
Efficient Distance Vector Routing for Mobile Ad Hoc Networks. In {\em Proceedings
of the 4th IEEE Workshop on Mobile Computing Systems and Applications (WMCSA 2002)},
IEEE, Calicoon, NY. June 2002.

\bibitem{HPJ01} Y. Hu, A. Perrig and D.B. Johnson, Ariadne: A Secure
On-Demand Routing Protocol for Ad Hoc Networks. {\em Proceedings of the Eighth Annual
International Conference on Mobile Computing and Networking (Mobicom 2002)}, pp. 12-23,
ACM, Atlanta, GA, September 2002.

\bibitem{HBC01} J. Hubaux, L. Buttyan and S. Capkun. The Quest for Security in Mobile
Ad Hoc Networks. {\em Proceeding of the ACM Symposium on Mobile Ad Hoc Networking
and Computing (MobiHOC)}, 2001.

\bibitem{BHcellular} M. Jakobsson, J.P. Hubaux, and L.
Buttyan. A Micro-Payment Scheme Encouraging Collaboration in Multi-Hop
Cellular Networks. {\em Proceedings of Financial Crypto 2003.}

\bibitem{DSR} D.B. Johnson and D.A. Maltz. Dynamic Source Routing in
Ad Hoc Wireless Networks. In {\em Mobile Computing}, edited by Tomasz
Imielinski and Hank Korth, Chapter 5, pages 153-181, Kluwer Academic
Publishers, 1996.

\bibitem{KZL01} J. Kong, P. Zerfos, H. Luo, S. Lu, and
L. Zhang. Providing Robust and Ubiquitous Security Support for Mobile
Ad Hoc Networks. In {\em International Conference on Network Protocols (ICNP)},
pages 251-260, 2001.

\bibitem{KL03} K. Lai, M. Feldman, I. Stoica, J. Chuang. Incentives
for Cooperation in Peer-to-Peer Networks. In {\em Workshop on Economics
of Peer-to-Peer Systems}, Berkeley, 2003.

\bibitem{MGLB00} S. Marti, T.J. Giuli, K. Lai, and M. Baker. Mitigating
Routing Misbehavior in Mobile Ad Hoc Networks. In {\em Proceedings of MOBICOM
2000}, pages 255-265, 2000.

\bibitem{PH02} P. Papadimitratos and Z.J. Haas. Secure Routing for Mobile
Ad Hoc Networks. In {\em SCS Communication Networks and Distributed Systems
Modeling and Simulation Conference (CNDS 2002), San Antonio, TX}, January 2002.

\bibitem{TESLA1} A. Perrig, R. Canetti, J.D. Tygar and D. Song. Efficient
and Secure Source Authentication for Multicast. In {\em Network and Distributed
System Security Symposium, NDSS '01}, Feb 2001.

\bibitem{TESLA2} A. Perrig, R. Canetti, J.D. Tygar and D. Song. Efficient
Authentication and Signing of Multicast Streams over Lossy Channels. In
{\em IEEE Symposium on Security and Privacy}, May 2000.

\bibitem{Roy02} K. Sanzgiri, B. Dahill, B.N. Levine, C. Shields,
E. Belding-Royer. A Secure Routing Protocol for Ad Hoc Networks. In {\em
International Conference on Network Protocols (ICNP)}, Paris, France, Nov. 2002.

\bibitem{SA99} F. Stajano and R. Anderson. The Resurrecting Duckling. Lecture
Notes in Computer Science, Springer-Verlag, 1999.

\bibitem{MultipathRouting} A. Tsirigos, Z. Haas. Multipath Routing in Mobile
Ad Hoc Networks or How to Route in the Presence of Topology Changes. In
{\em Proceedings of IEEE MILCOM 2001}.

\bibitem{GL98} X. Zeng, R. Bagrodia, M. Gerla. GloMoSim: A Library
for Parallel Simulations of Large-Scale Wireless Networks. {\em Proceedings
if the 12th Workshop on Parallel and Distributed Simulations -- PADS '98},
May, 98, Alberta, Canada.

\bibitem{ZH99} L. Zhou and Z. Haas. Securing Ad Hoc Networks. In
{\em IEEE Network magazine, special issue on networking security, Vol. 13,
No. 6, November/December}, pages 24-30, 1999.

}
\end{thebibliography}

%

%

\end{document}